# Pattern Formation in Isothermal Miscible Protein/Sugar Systems Driven by Marangoni Effects and Evaporation

Yu-Ching Tseng[1‡], Chamika Goonetilleke[1‡], Xiaotian Lu[2‡], Niladri Sekhar Mandal[2], Ali Borhan[2*], Ayusman Sen[1,2*]

[1]Department of Chemistry, Pennsylvania State University, University Park, Pennsylvania, USA

[2]Department of Chemical Engineering, Pennsylvania State University, University Park, Pennsylvania, USA

Email: axb20@psu.edu, asen@psu.edu

[‡]Equal Contributions

**Abstract**

Through a combination of experiments and modeling, we have demonstrated a novel pattern formation phenomenon in an isothermal miscible fluid system involving simple protein and sugar solutions. We introduced dye-tagged protein solution into a petri dish with sugar solutions, which had higher density than the added protein solution. Initially, the protein spread and became more uniformly distributed at the air-water interface. Subsequently, it concentrated in specific areas to form spiral patterns. We propose that the mechanism involves an interplay between Marangoni effects, evaporation, and airflow. This finding is unexpected as solute Marangoni-related processes are generally characterized by fast spreading (seconds), while the pattern formation in our systems takes several minutes to form. Our work suggests that Turing reaction-diffusion patterns can be replicated by replacing the reaction-induced inhomogeneous solute distribution by evaporation-induced inhomogeneity. In both cases, the fast diffusive or Marangoni spreading of the solute is counteracted by a slower step that serves to reverse the solute homogenization. In showing that dissipative patterns can form in the absence of thermal gradients or chemical reactions, our findings significantly expand the conditions that lead to pattern formation. The insights gained also enhance our ability to manipulate and control fluid motion and surface morphology, with promising implications for many areas such as coating technologies, materials science, and microfluidics.

## Introduction

We show novel pattern formation in an isothermal miscible fluid system involving simple protein and sugar solutions. Nature is full of intricate and fascinating patterns, with complex underlying principles.[1,2] Beyond aesthetics, the spatiotemporal organization within living systems is crucial in multiple aspects, such as cell polarization[3], regulation of cell division[4], and behavior of bacteria in a dense suspension.[1,5] The reaction-diffusion mechanism has been widely accepted as a primary model for pattern formation.[2,6–8] However, mechanisms other than chemical reactions can also induce instabilities if they disrupt equilibrium conditions.[9] For example, Marangoni[10,11], Rayleigh–Bénard[12], and evaporation effects[13] are three physical processes that can also independently or collectively drive fluid motion and lead to pattern formation.

Marangoni effects occurs when there is a gradient in interfacial tension at the interface of two fluids, which can result from temperature gradients (thermocapillary effect) or concentration gradients (solutocapillary effect).[14] This effect drives fluid flow from regions of lower interfacial tension to higher interfacial tension. Consequently, the flow destabilizes the initially uniform conditions at the interface , leading to the emergence of various patterns such as waves, cells, or other structured formations.[15–17] Cartwright et al. reported patten formation due to the interaction between Marangoni and Rayleigh–Bénard effects, often referred to as Bénard–Marangoni convection.[18]

Evaporation from an aqueous surfactant solution can also play a significant role in Marangoni instability. As the interfacial concentration of a surface-active species changes non-uniformly due to local evaporation of water, an interfacial tension gradient becomes established. The resulting Marangoni effect can impact heat and mass transfer processes in droplets.[19]. For example, Van Gaalen et al. have shown that stronger evaporation can lower the solubility of surfactants, leading to suppression of Marangoni circulation in evaporating sessile droplets.[20] The interaction between the Marangoni effect and evaporation can lead to pattern formation observed in both experiments and simulations.[21–24] For example, Wodlei et al. investigated the complex dynamics involved in the evaporation of liquid drops on an immiscible aqueous substrate, particularly focusing on the formation of unique flower-like patterns driven by evaporation-induced Marangoni flow.[23]

Most of the reported systems involve *immiscible* fluids, or patterns formed in thin films or on solid plates.[21,25] For these cases, the interphase interactions can significantly influence pattern formation.[23,25] In contrast, pattern formation in miscible fluid systems remains poorly characterized and typically involves thermal gradients.[26–28] When two or more miscible fluids with different properties are combined, they do not simply mix; they can interact in complex ways driven by differences in density, concentration, temperature, or surface tension. For instance, Strombom et al. observed pattern formation when high-density food dyes were added to a shallow dish of water solution.[29] However, their use of a Turing-like model is not appropriate due to fundamental differences in the underlying mechanisms and system behavior.

In this study, we report a novel pattern formation behavior in a miscible protein/sugar fluid system over a longer mixing period (mins) (**Figure 1**) under isothermal conditions. We introduced a dye-tagged protein solution at the air-liquid interface in a petri dish containing sugar solutions with higher density than the added protein solution. Initially, the protein spread and became more uniformly distributed at the air-water interface. Subsequently, it concentrated in specific areas to form spiral patterns. We hypothesize that the underlying mechanism involves an interplay between Marangoni effects, evaporation, and ambient airflow. We developed computational models to validate our hypothesis and reproduce the observed patterns. We also conducted additional experiments to further examine the mechanisms suggested by the simulations. We discuss how the interplay between evaporation and Marangoni effects leads to pattern formation in miscible fluid systems. Our findings are unexpected in that fast Marangoni flow-induced spreading and homogenization[26,30,31] *precedes* the subsequent slower pattern formation, which takes several minutes to develop due to the competition between evaporation and reactivated weaker Marangoni effects. Thus, the system undergoes spontaneous symmetry breaking and transition between equilibrium and non-equilibrium states, ultimately generating dissipative patterns over time.

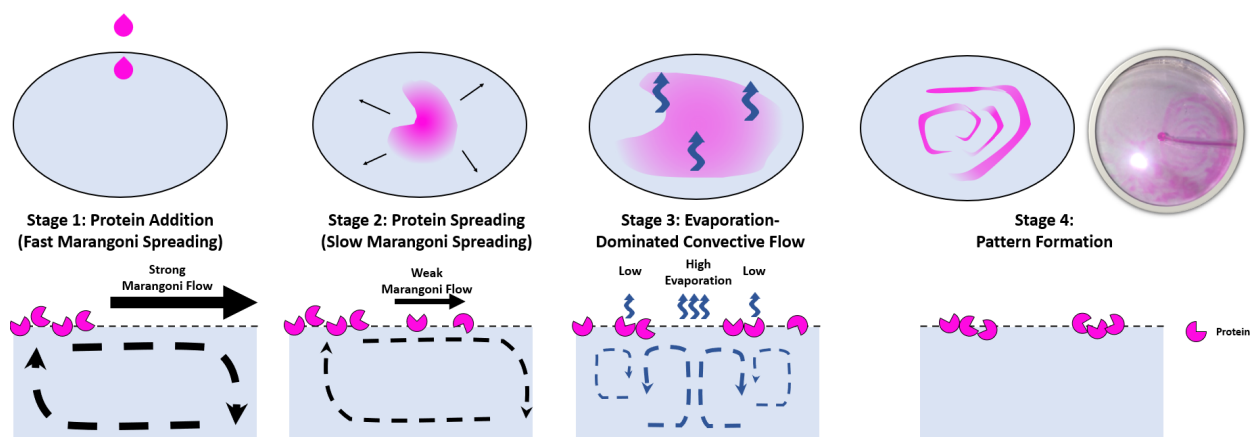

**Figure 1.** (A) Schematic illustration of pattern formation of dye-tagged proteins at air/water interface. After addition, the density difference between protein solution and glucose solution keeps the protein solution at the interface. The surface tension drops due to the presence of the protein and asymmetric local protein concentration leads to a Marangoni flow, which causes the protein to spread and become more uniformly distributed. While the Marangoni flow weakens over time, evaporation causes surface mass and momentum loss, prompting the bulk solution with low protein concentration to convectively transport to the surface, increasing the local protein concentration difference. As a result, protein patterns form. The images in the top row show the pattern formation process from the top view and the images in the bottom row show the process from a side view.

## Results and Discussion

**Pattern formation with urease**

In our experimental setup, as shown in **Figure S1**, a syringe containing a dye-tagged protein solution was installed onto a syringe pump and connected to a tube. The end of this tube was positioned approximately 0.5 cm above the water level within a polystyrene petri dish. Then 50 μl of 50 μM Dylight 550-tagged urease, suspended in 200 mM MES buffer, was introduced into the petri dish containing 100 mM D-glucose in 200 mM MES buffer. The entire process was recorded by a camera and a sample video is provided in **Supporting Video S1**. Snapshots from the video are presented in **Figure 2A** to highlight significant events in the pattern formation process. During the addition, multiple droplets of dye-tagged urease were dropped into the petri dish, quickly spreading at the air-water interface, and creating a nonuniform protein distribution. After one minute, Marangoni flow induced a relatively homogeneous distribution of protein. Subsequently, urease started to concentrate into specific areas and formed a spiral pattern, which remained visible for up to five minutes. In contrast, spiral pattern formation was not observed and urease slowly spread (in the upper left region of **Figure 2B**) when a similar experiment was conducted by adding dye-tagged urease into 200 mM MES buffer *without* D-glucose (**Figure 2B and Video S2**).

To gain further insight into pattern formation, we analyzed pixel intensity profiles from **Figures 2A** and **2B** at specific times after addition (3 and 5 min). The resulting profiles are displayed in **Figures 2C and 2D**, and the locations of the analyzed lines are indicated in the insets of each figure. As shown in **Figure 2C**, when the pattern begins to form, pronounced fluctuations can be seen in the intensity profiles at 3 and 5 minutes, suggesting significant changes in urease distribution in the petri dish. Conversely, in the group where only MES buffer was present in the petri dish (**Figures 2D**), the intensity profiles at 3 minutes and 5 minutes remain essentially flat. This observation reflects a stable and uniform distribution of urease in the absence of glucose.

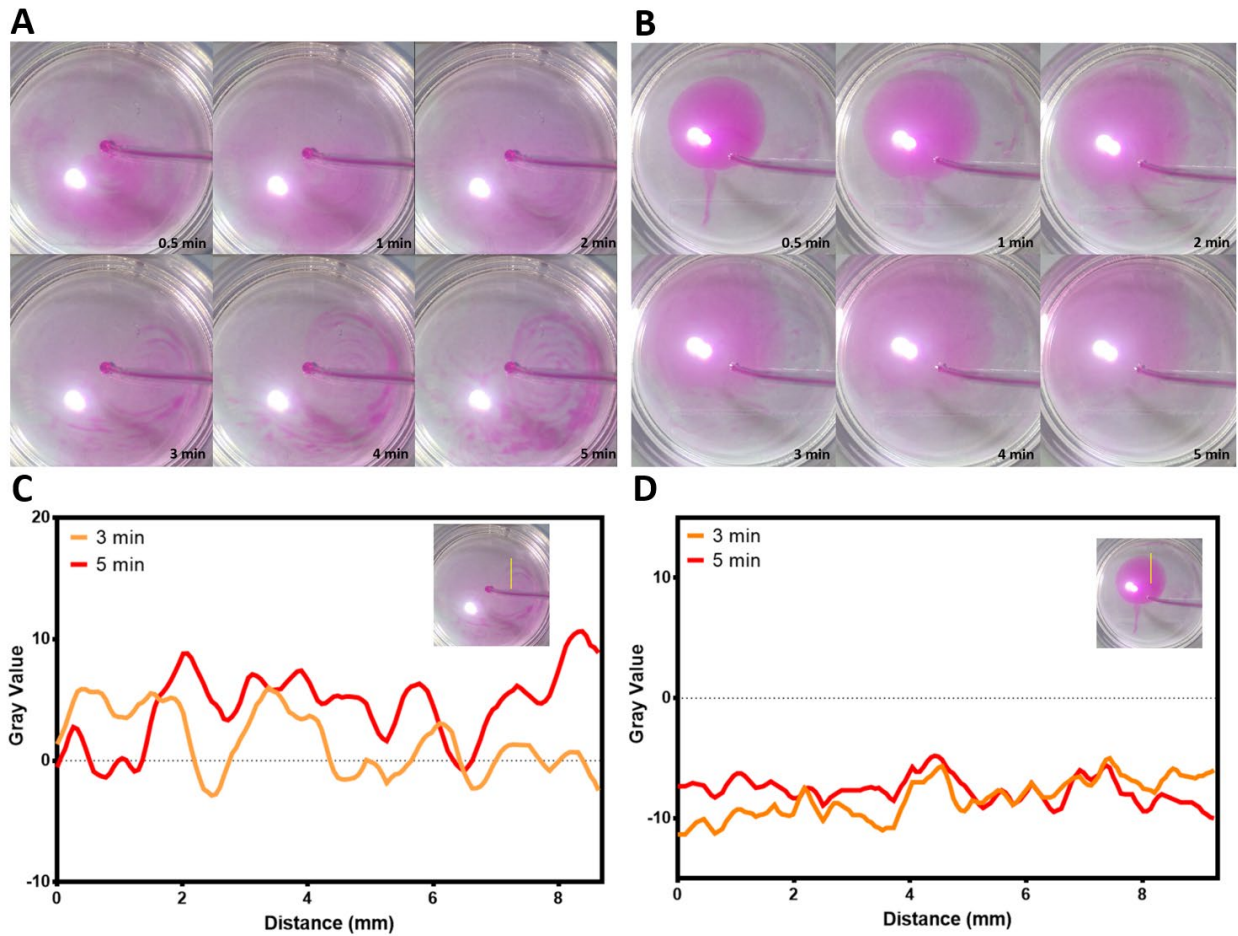

**Figure 2.** (A-B) Snapshots of Dylight550-tagged urease in (A) 200 mM MES buffer with 100 mM glucose and (B) 200 mM MES buffer only. All solutions were prepared in 200 mM MES buffer, pH 6. (C-D) Pixel intensity along line as shown in MES buffer with 100 mM glucose and MES buffer only. The locations of the lines analyzed are indicated in the image insets. Background intensity before protein addition was subtracted from the intensity profiles at each time point.

The same experiments were conducted by introducing urease into either MES buffer alone or MES buffer containing 100 mM glucose, with the process being monitored from the side using a camera. This perspective provides a clear observation of the dye-tagged protein's location during pattern formation. When urease was added into the 200 mM MES buffer containing 100 mM D-glucose, it was observed to float on the surface throughout the pattern formation process (**Figure S2 and Video S3**). During the first two minutes after addition, urease floated and spread at the air-water interface, with minimal movement detected. However, noticeable underwater movement of the protein was observed approximately 2.5 minutes post-addition when the pattern began to form. On the other hand, in experiments using MES buffer without glucose in the petri dish, urease was seen to sink directly to the bottom and spread out across the bottom of the petri dish (**Video S4**).

From these side-view observations, we can designate pattern formation in these experiments as a surface phenomenon. Consequently, we hypothesize that the presence of the protein at the air-water interface is crucial for pattern formation. This phenomenon occurs due to the density difference between the urease solution and the solution within the petri dish. Specifically, the addition of protein suspended in a less dense solution into a denser solution within the petri dish enables the protein to float at the air-water interface, where pattern forms.

To validate the proposed hypothesis, urease was dissolved in buffer with 100 mM glucose and introduced into various solutions: 200 mM MES buffer, and MES buffers containing 100 mM glucose or sucrose. Considering the densities of solution, buffer with 100 mM sucrose has the highest density, followed by buffer with 100 mM D-glucose, and then 200 mM MES buffer alone (**Figure S3**). We hypothesized that urease in 100 mM glucose would only form patterns when it was added to the denser 100 mM sucrose solution in buffer. The experimental results validated our hypothesis. As depicted in **Figures S4A and S4B**, urease in both MES buffer and glucose solution spread uniformly without forming patterns. In contrast, urease in the sucrose solution started to form pattern approximately three minutes after adding protein to the sucrose solution (**Figure S4C**). These experiments clearly demonstrate that pattern formation happened when proteins suspended in a less dense solution were introduced into a denser solution. The videos for glucose and sucrose experiments are shown in **Videos S5 and S6**, respectively.

To further investigate the underwater movement of the protein in the later stages, fluorescence microscopy was employed. In this experiment, the focal plane was set to 0.5 mm below the air-water interface. When introducing urease into MES buffer containing 100 mM glucose, underwater movement was observed approximately 2.5 minutes after the addition (**Figure S5A and Video S7**), aligning with the pattern formation timing previously observed from top and side views using the camera. Conversely, with MES buffer alone in the petri dish, no significant movement was observed (**Figure S5B and Video S8**).

Particle image velocimetry (PIV) analysis was applied to video recordings to investigate the movement of urease and its association with pattern formation. The detailed methodology of this analysis is provided in the **Materials and Methodology** section. In brief, the PIV analysis was conducted every second and the average speed of urease was calculated at 30-second intervals after its addition into the petri dish. To visually represent the movement of urease at different locations, we generated color maps of the average speed for two critical time intervals: 61-90 seconds and 151-180 seconds (**Figures 3A to 3D**). These intervals were specifically selected due to their significance in the pattern formation process. The first interval, 61-90 seconds, was chosen to ensure that any disturbances caused by the initial introduction of urease had disappeared, thereby providing a baseline speed of urease prior to any pattern formation. The second interval, 151-180 seconds, corresponds to the start of pattern formation, typically observed approximately two to three minutes after adding the protein.

The color maps depicting the speed of urease in 200 mM MES buffer with 100 mM glucose for the time intervals of 61-90 seconds and 151-180 seconds are shown in **Figures 3A and 3B**, respectively. Similarly, **Figures 3C and 3D** depict the speed color maps of urease in 200 mM MES buffer without glucose for the same time intervals. By comparing **Figures 3A and 3B**, multiple regions with higher speed can be observed during the 151-180 second interval. Notably, an area in the top right of **Figure 3B** shows a velocity peak reaching 50 µm/s. These regions of increased speed, when compared with **Figure 1B**, correspond to the locations where patterns form. This suggests urease molecules move to concentrate in specific areas, leading to pattern formation. In contrast, the speed of urease in 200 mM MES buffer is lower and tends to decrease over time, as illustrated in **Figures 3C and 3D.**

To further understand how the speeds evolve over time, speed distributions during the chosen time interval (61-90 seconds and 151-180 seconds) are plotted with the speed on the X-axis and percentage of vectors on the Y-axis (**Figure 3E**). For this analysis, only vectors with speeds exceeding 5 µm/s were included. The percentage of vectors within a given range of speed was determined by counting the number of vectors within each 1 µm/s speed interval (For example, 5-6 µm/s, 6-7 µm/s, et al.), then dividing that count by the total number of vectors with speeds over 5 µm/s during the chosen time period. In the first time interval (61-90 seconds), the speed distributions were similar in 100 mM glucose and MES buffer. However, a notable increase in speed in the 15 to 30 µm/sec range was observed in 100 mM glucose during the 151-180 seconds interval. This enhancement occurs at the onset of pattern formation (**Figure 1A**). In contrast, in the MES buffer, the speed decreased over the same time period. Using the same method to analyze fluorescence microscopy videos led to qualitatively similar results for underwater movement of protein. Namely, as illustrated in **Figure S6,** urease in glucose showed an increase in speed during the 211-240 second interval compared to the 61-90 second interval, whereas urease in MES buffer barely showed an increase.

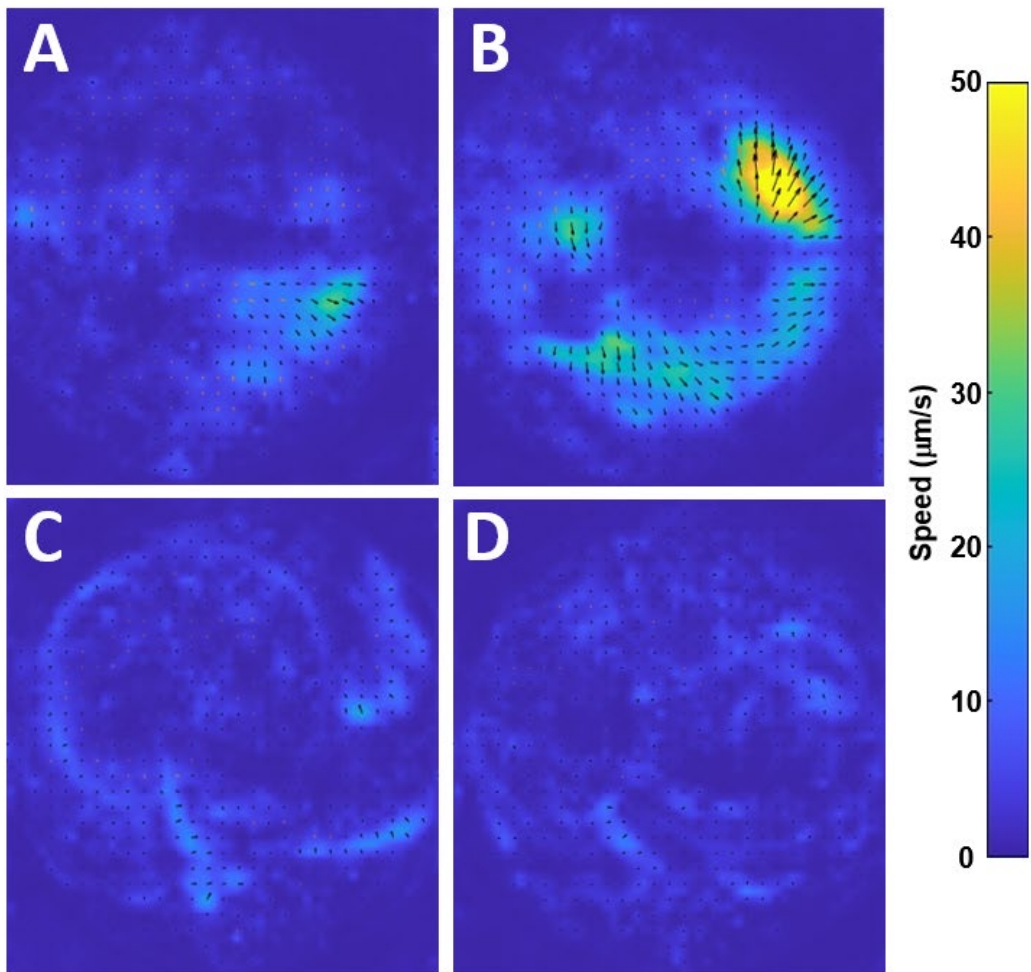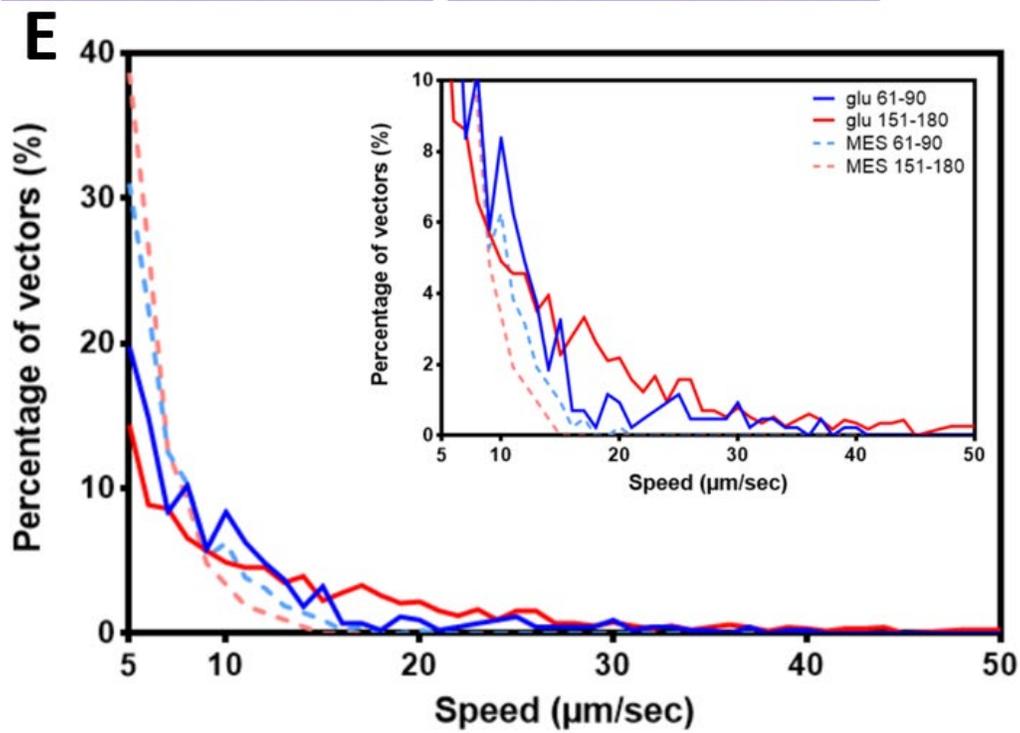

**Figure 3**. (A-B) Speed color maps of urease in 200 mM MES buffer with 100 mM glucose. (A) Average speed map of urease from 61 to 90 seconds after urease addition. (B) Average speed map of urease from 151 to 180 seconds after urease addition. (C-D) Speed color maps of urease in 200 mM MES buffer only. (C) Average speed map of urease from 61 to 90 seconds after urease addition. (D) Average speed map of urease from 151 to 180 seconds after urease addition. Black arrows represent the velocity vector based on analysis and orange arrows represent the vector that is interpolated based on the nearby vectors. (E) Speed distribution of urease in the selected time interval. Only vectors with speeds greater than 5 µm/s were considered. The percentage of vectors was calculated by dividing the number of vectors in each 1 µm/s speed interval by the total count of vectors above 5 µm/s in the selected time interval. The inset shows the percentage range from 0 to 10%.

**Numerical Simulation for Mechanism Verification**

Based on the experimental observations described earlier, we categorize the process of pattern formation into four stages, as shown in **Figures 1 and 4**. The observed pattern formation at the air-water interface can be attributed to interactions among solutal Marangoni flow, evaporation, and shear-induced convection by ambient air flow. In stage one, the added urease solution acts as a surfactant, locally lowering the surface tension and inducing a strong Marangoni flow. Transitioning to stage two, the flow weakens over time due to the decrease in the concentration gradient of urease at the interface; eventually, a nearly uniform distribution of protein is reached. In stage three, the effect of evaporation becomes dominant over the Marangoni effect. Evaporation causes surface mass and momentum loss, generating convection loops that mix the bulk with the surface layer (a process known as evaporative convection)[13]. At this stage, slight differences in protein concentration still exist at the interface. The regions with higher protein concentrations have lower evaporation rates, and those with lower concentrations have higher evaporation rates. Convective flows are generated due to the difference in evaporation rates, causing proteins in lower concentration regions to move to areas of higher concentrations, amplifying the local protein concentration differences. This, in turn, creates a new surface tension gradient, reactivating the Marangoni effect and resulting in a competition between Marangoni flow and evaporative convection. As a result, protein patterns appear in stage four. The shear-induced surface convection resulting from ambient airflow can further influence the surface redistribution of urease, creating different types of visual patterns.

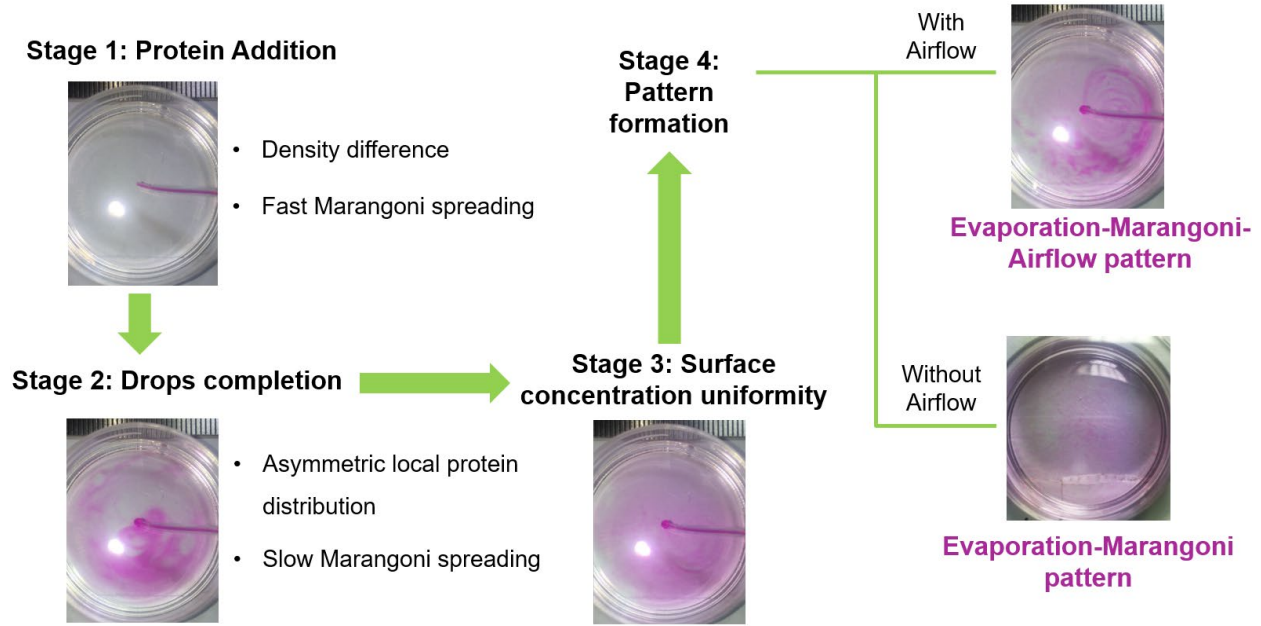

**Figure 4**. Categorized stages for pattern formation in the miscible urease/glucose solution system. Stage 1: When low-density urease solution is dropped into the glucose solution, the initial density difference causes the urease solution to float at the water-air interface, and the local surface tension difference induces strong local Marangoni flows. Stage 2: Approximately 10 seconds after the dropping process, an asymmetric protein distribution forms at the surface, characterized by a slower Marangoni spreading due to reduced surface tension gradient compared to stage 1. Stage 3: A relatively uniform urease concentration develops as Marangoni flow minimizes surface concentration differences. With the fading of the Marangoni effect, evaporation begins to play a more dominant role. Stage 4: Evaporative convection loops mix the bulk and surface layers, increasing surface protein concentration differences and reactivating the Marangoni effect. Patterns emerge due to the competition between evaporation and the Marangoni effect. Additionally, airflow-induced surface perturbations influence the types of patterns: without a lid, spiral patterns form, while covering the petri dish with a lid to eliminate airflow results in cell-like patterns.

Computational fluid dynamics (CFD) models using Ansys Fluent software were developed to simulate the different stages of pattern formation, verify the proposed mechanisms, and reproduce the observed pattern (See **Supported Information** and **Table S1** for more details). The species transport model was chosen to represent the experimental miscible fluid system of urease/glucose solution, and transient simulations were performed to capture the dynamic pattern formation. To study the different scenarios of pattern formation, modules for solutal Marangoni stress, surface evaporation, and airflow-induced surface shear stress were developed in Ansys Fluent (See **Materials and Methodology** and **Supporting Information** for more information). In addition, to balance computational cost and efficiency for three dimensional (3D) transient simulations, multiple models were employed to investigate key factors at each stage separately.

In our system, we identified a two-stage solutal Marangoni spreading process for protein solution. Solutal Marangoni flow is typically characterized by high surface velocity and can quickly eliminate surface concentration gradients in small-scale systems. Therefore, studies involving solutal Marangoni flow in miscible systems often focus on short time scales.[26,30] However, our experimental system achieved a relatively uniform surface concentration at around ~40-50 sec, which is significantly longer than reported time scales in the literature (**Video S1**). As shown in **Video S1**, we dropped multiple droplets of 50 µM buffered urease solution (~12 drops, total volume of 50 µL) into a petri dish containing 100 mM buffered glucose solution. Fast Marangoni spreading and surface solute transport occurred within seconds of protein addition. By balancing Marangoni stress with viscous drag, we estimated the Marangoni velocity (>0.06 m/s) and the timescale (<0.3 s) for spreading of a single protein droplet (**Supporting Information, Section S2**). This aligns with the fast transient spreading and Marangoni instabilities observed in **Video S1** as each protein drop is introduced. However, in this miscible system, achieving a relatively uniform surface concentration took considerably longer (~1 minute, as shown in **Video S1**). This extended time can be attributed to the difference between adding a single drop and multiple drops. Protein droplets added later are affected by the concentration field established by earlier drops. The weaker overall Marangoni driving force ($\partial\sigma/\partial C \downarrow, \nabla C \downarrow$) and the counteracting local Marangoni flows slow the protein spreading process (stage 2), thereby increasing the time required to achieve a uniform surface concentration.

Based on the above consideration, we investigated the effect of different surface protein patches in stage two (slow Marangoni spreading process) through CFD simulations (see **Supporting Information, Section S3** for more details). Urease patches with two different initial configurations were considered: a radially symmetric urease patch (**Figure S7A**) and an asymmetric urease patch (**Figure S7B**) emerging from the previous spreading stage (based on our experiments), corresponding to ideal and non-ideal droplet injection and spreading processes, respectively. According to the simulation results, the symmetrical urease patch (**Figure S7A**) spreads isotropically with a velocity that is dependent on the instantaneous patch radius. In contrast, the asymmetric urease patch exhibits an irregular spreading profile due to the asymmetric initial surface protein concentration and Marangoni stress distributions (**Figure S7B**). In addition, an asymmetric protein distribution can generate opposing local Marangoni flows that counteract each other. Therefore, the asymmetric urease patch experiences a longer Marangoni spreading process.

**Figure 5** shows the surface concentration distribution for the asymmetric urease patch at different spreading times. As shown, the asymmetric distribution of Marangoni stress drives an asymmetric surface flow, with the surface concentration variation decreasing to less than 1 µM after 32 seconds (also see **Video S9** for the simulation up to 50 seconds). This aligns with the time scale in our experiments required to achieve a relatively uniform surface concentration (**Video S1**). Based on these simulations, an asymmetric surface protein distribution formed after stage one is essential for pattern formation. Without this nonisotropic concentration distribution, surface flow and solute transport would remain radially symmetric, preventing pattern formation at later stages.

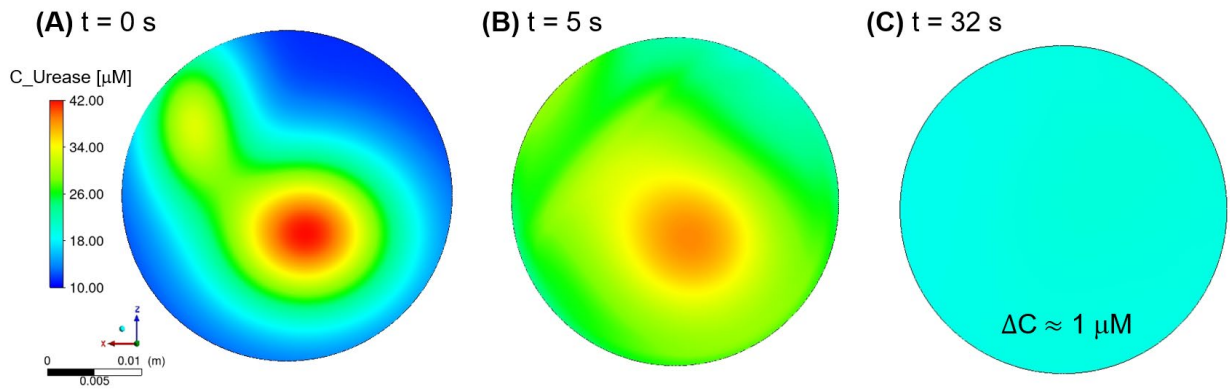

**Figure 5**. Simulation of the slow Marangoni spreading stage (stage two) for an asymmetric surface urease patch (top view). The asymmetric concentration gradient leads to asymmetric Marangoni flows, driving preferential surface movement. The surface urease concentrations at different times are displayed: (A) t = 0 s, (B) t = 5 s, and (C) t = 32 s, where a relatively uniform surface concentration is achieved (see **Figure S8** for urease local concentration distribution when $\Delta C \approx$ 1 µM and **Video S9** for the complete 50-sec simulation).

Pattern formation in the third and fourth stages results from the interaction between Marangoni, evaporation, and airflow effects. The evaporation at the surface results in compensatory fluid flow from the bulk to the surface. This flow velocity, estimated from the evaporation rate ($v_{eva} = \frac{Q_{eva}[m^3/s]}{A_{petri\ dish}[m^2]}$), is many orders of magnitude lower than the Marangoni spreading velocity during the first two stages. For example, an average evaporation rate of 1 × $10^{-4}$ kg/(m²·s) yields an estimated flow velocity of 0.1 µm/s (see **Supporting Information, Section S1** for more details). Therefore, during the earlier spreading stages, when the stronger Marangoni flow dominates, evaporation-induced fluid flow can be discounted. As the surface tension gradient and Marangoni velocity diminish, evaporation begins to play a more prominent role in mass transport.

In stage three, the system starts with a relatively uniform surface protein concentration (**Figures 1 and 4**). Evaporation causes fluid mass and momentum loss near the surface. As a result, bulk fluid is transported to the surface, thereby creating convective cells in the bulk.[13] Additionally, since the saturation pressure of water is lowered by the presence of the protein, the water evaporation rate varies locally at the surface due to spatial variations in protein concentration,. As a result, multiple convective cells with varying magnitudes form within the bulk fluid, leading to localized concentration or dilution of the protein at the surface. This drives the formation of patterns.

In the simulation of stage three, an asymmetric urease patch model with a surface concentration difference of about 1 µM was developed to study this evaporative-convection effect. Our simulation predicts the formation of irregular cell-like patterns (**Figure 6A and Video S10**), driven by evaporation as well as the accompanying underwater dynamics. In **Figure 6B**, the vertical velocity distribution reveals the presence of convective cells and momentum transfer between the bulk and the surface. These convective cells lead to the irregular surface velocity distribution shown in **Figure 6C**. In addition, it was found that reducing the water evaporation rate would increase the time required for pattern formation and reduce the concentration difference on the surface (**Figure S9**). These simulation results well explain the impact of evaporative convection and support our proposed mechanism for pattern formation. The underwater flows predicted by our simulation align with our experimental observations. For example, enhanced underwater movement of the protein was observed after 2-3 min in side-view videos as patterns began forming (**Figure S2** and **Video S3**) or through a fluorescence microscope (**Figure S5** and **Video S7**). The velocity characterized from experiments when the pattern formed (**Figure 3E**) is also of comparable magnitude to the simulated velocity shown in **Figure 6C**.

The pattern (irregular cells) in **Figure 6A** is different from those observed in experiments (**Figure 2A**). Since our experiments were conducted in an open petri dish, the pattern formed at the air-water interface can be affected by the surrounding environment, such as ambient air currents. Airflow can potentially increase the evaporation rate by decreasing surface humidity, and exert shear stress at the air-water interface to alter the surface flow.[32–34] To explain the experimental observations, we considered a more realistic scenario, where the subtle surface disturbances caused by airflow were modeled by applying weak shear stress to the air-water surface. Two types of airflow scenarios were considered given the rotational surface flow observed within the cavity of the petri dish during experiments: symmetric and asymmetric shear stress. These scenarios triggered an average surface flow speed of 9.5 µm/s (**Figure S10A**) and 1.05 µm/s (**Figure S10C**), respectively, based on the simulation. With the addition of airflow disturbance, the pattern transitioned from cell-like to spiral forms (**Figure 6D** and **Video S11**). Ultimately, we successfully reproduced the spiral pattern observed in our experiments (**Figure 2A**). A similar spiral pattern was also observed for the asymmetric airflow (**Figure S10D**).

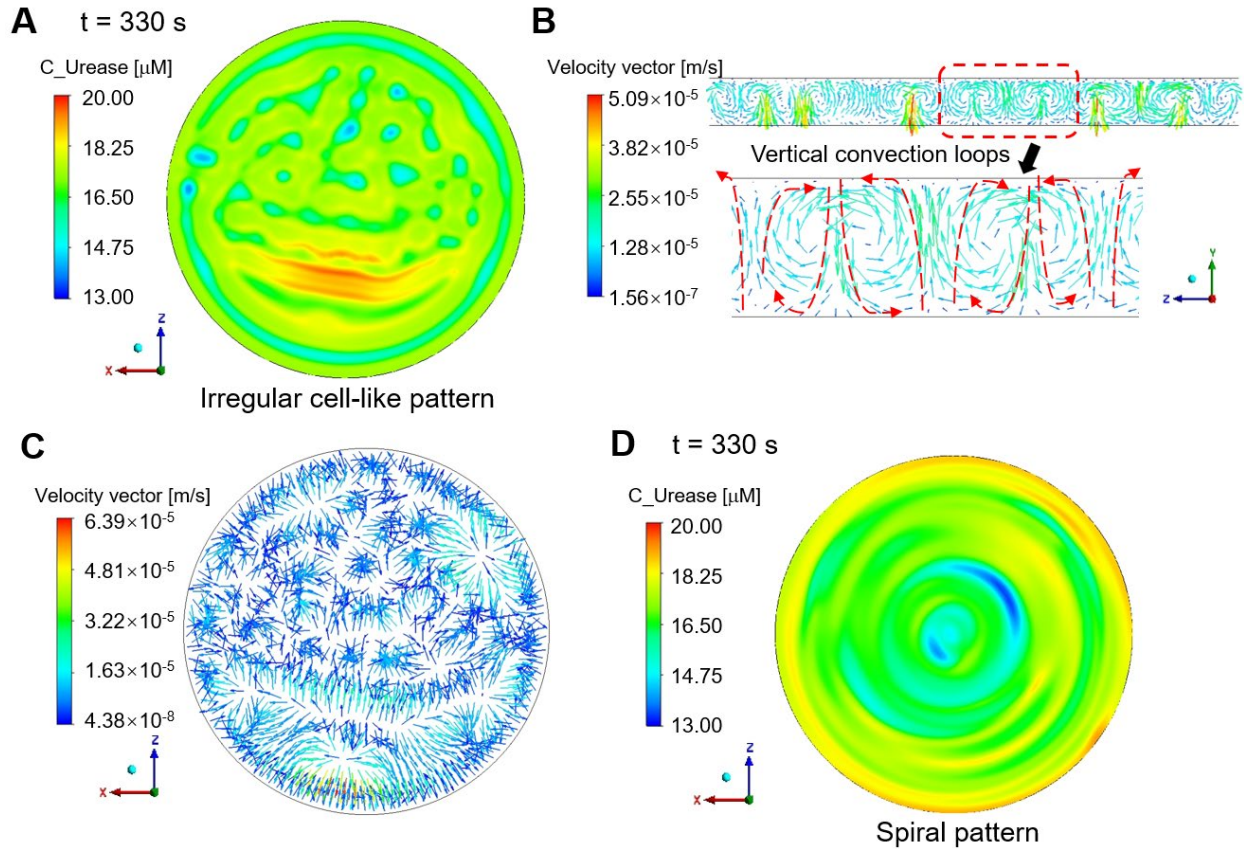

**Figure 6**. Simulated (A-C) irregular cell-like surface pattern with no surrounding airflow influence at a simulation time of 330 s: (A) urease concentration contour, (B) vertical plane velocity vector illustrating underwater convection loops (Red dash lines were added to guide the eye), and (C) surface velocity vector exhibiting complex local flow behaviors, leading to local concentration and dilution of urease solution (Velocity vector arrows were normalized for a better visualization). When considering airflow effect, surface pattern shifts from an irregular-cell like to (D) a spiral form.

To investigate the effect of airflow on pattern formation experimentally, we conducted two separate experiments with different environmental airflow settings. In the first experiment, a fan was positioned near the petri dish, with strong airflow above the solution surface during the experiment. This caused a noticeable, unified single rotational flow pattern and rapid mixing of species (**Figure S11; see Video S12 for clear dynamic rotational speed**). This result confirms that airflow can drive surface flow and induce rotational movement. However, since the airflow from the fan was significantly stronger than typical lab ventilation, no pattern was observed in this case due to fast mixing. Using simulations, we could reduce the airflow velocity to mimic a more realistic environment in the laboratory. Using the perturbation caused by airflow that was sufficient to induce surface movement but without accelerating species mixing, we successfully simulated the spiral pattern (**Figure 6D**). These findings indicate that pattern formation is highly sensitive to even small system perturbations.

In the second experiment, the petri dish was covered with a transparent lid to eliminate the airflow perturbation from the laboratory environment (**Video S13**). Snapshots from the video are presented in **Figure 7.** Our experiment successfully replicated the formation of irregular cell-like patterns predicted by the simulations (**Figure 6A**). In addition, the pattern generated when the petri dish was covered by a lid was less evident compared to an open petri dish (**Figure 2A**). The time required for pattern formation was also increased, taking 3-4 min with the lid on compared to 2-3 min without it. This can be attributed to the reduction in evaporative flux, as confirmed by our simulations (**Figure S9**).

These experimental results, combined with the surface velocity characterization (**Figure 3**) and the underwater dynamics observed from the side-view (**Figure S2 and Video S3**) or using fluorescence microscope (**Figure S6 and Video S7**), provide solid experimental evidence to support our proposed mechanism for pattern formation.

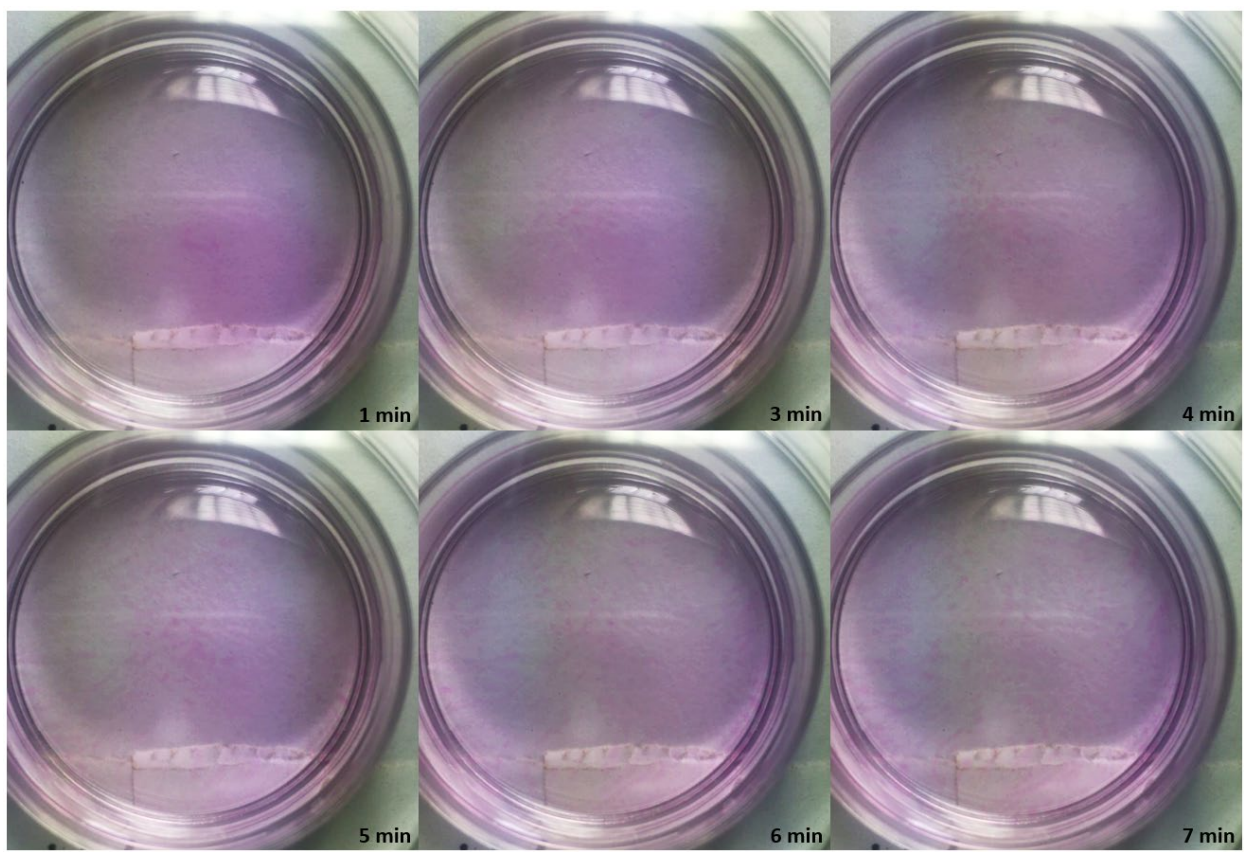

**Figure 7**. Snapshots of pattern evolution of Dylight550-tagged urease added to 200 mM MES buffer with 100 mM glucose. The petri dish was covered with a transparent lid to eliminate the airflow effect on pattern formation. An irregular cell-like pattern was observed. All solutions were prepared in 200 mM MES buffer, pH 6.

As previously discussed, protein can act as a surfactant due to its amphiphilicity, thereby reducing the surface tension of aqueous solutions. To investigate the influence of enzyme amphiphilicity on Marangoni flow during pattern formation, 100 µM (in place of usual 50 µM) dye-tagged urease in buffer was introduced into a petri dish with buffered 100 mM glucose solution. To consistently compare different groups, we defined "pattern formation time" as the post-addition time required for protein to transit from a relatively homogeneous state to form a pattern. In other words, this is the point at which any recognizable pattern, such as strips, first becomes apparent. In **Figure 8A**, the time required for pattern formation was shorter when adding 100 µM urease compared to 50 µM. With increased protein concentration, a faster and stronger Marangoni spreading process was observed due to a greater local surface tension gradient. This accelerated the first two stages of pattern formation, reducing the overall time needed for pattern formation.

In the previous section, we demonstrated the necessity of a density difference for pattern formation. To determine the minimal density difference required to form patterns, tagged urease suspended in 200 mM MES buffer was added into the same buffer with 20, 40, 60, 80, 100, and 200 mM glucose. We found that the pattern stopped forming for the MES buffer with 20 mM glucose solution, suggesting that the minimal density difference is approximately 0.00256 g/cm$^3$ This value was deduced based on the density difference between buffer solutions of 0 and 40 mM glucose, as presented in **Figure S3**.

The time taken to form patterns is dependent on the concentration of sugar solution in the petri dish. In **Figure 8B**, we plot the pattern formation time versus concentration of glucose solution in millimolar. Notably, a linear relationship was observed between the time of pattern formation and the concentration of glucose. To further confirm the linear relationship, we introduced tagged urease into 200 mM MES with varying concentrations of sucrose, ranging from 20 to 100 mM. An analogous linear relationship but with a higher slope was observed (**Figure 8B**).

We plotted the pattern formation time against the density of solution in **Figure S12**, showing similar trends for the two different solutions. This suggests that the observed pattern formation times are closely related to the density changes in the solution, within a specific density difference range—from 0.00256 g/cm$^3$ to 0.0118 g/cm$^3$. For lower density differences, less protein remains on the free surface due to the reduced buoyancy effect. This reduces the surface protein concentration, resulting in weaker Marangoni flow, allowing evaporative convection to dominate more quickly, which accelerates pattern formation. As the density difference increases, a stronger Marangoni flow is achieved, extending the time required for pattern formation. For very low-density differences, no pattern forms as the protein does not remain on the free surface.

Another factor affecting this relationship is the vapor pressure of the sugar solution. Higher sugar concentration can decrease the vapor pressure of solution and slow down the evaporation process. In **Figures 7** and **S9**, we have shown that with lower evaporation flux in the system, the pattern formation time becomes longer.

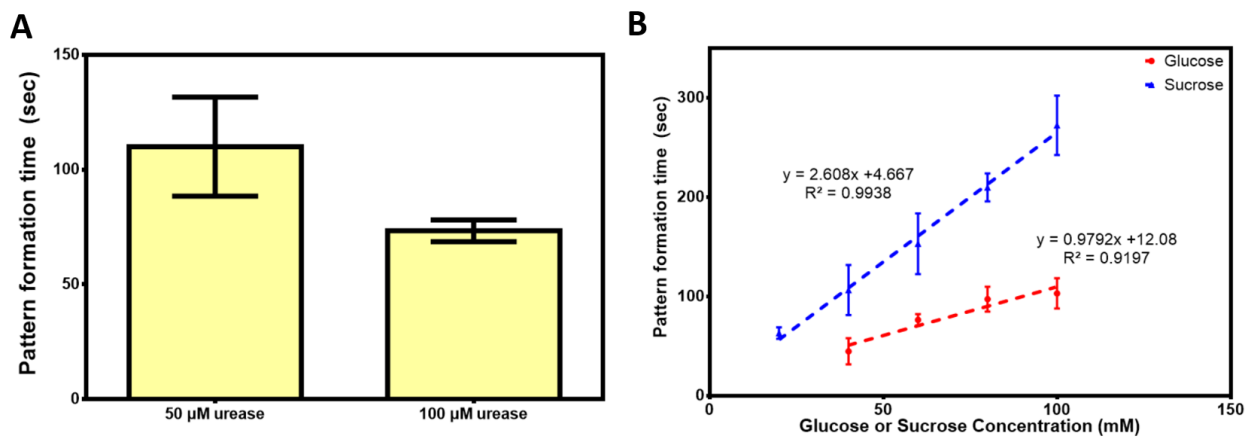

**Figure 8.** (A) Time required for different concentrations of urease to start forming pattern after being introduced into petri dish containing 100 mM glucose. (B) Linear relationship between pattern formation time of urease and the concentration of glucose or sucrose solutions in the petri dish. All solutions were prepared in 200 mM MES buffer, pH 6. The error bars are derived from 3 different trials.

The results of this study offer new insights into the mechanism underlying pattern formation in miscible solution systems under isothermal conditions by identifying the interplay between evaporation and solutal Marangoni convection as the key contributing factor. As the former acts to amplify concentration differences via water loss from the surface and replenishment from the bulk phase, the latter acts to diminish existing concentration differences.

In addition, even a small perturbation (such as that caused by surrounding air flow) might be amplified in the system to significantly affect the eventual pattern. Using our computational models, we conducted a sensitivity analysis to examine how variations in Marangoni and evaporation effects affect the pattern formation results. A phase diagram was developed (**Figure 9**) based on those results, illustrating the interplay between Marangoni and evaporation effects on surface pattern formation, keeping the airflow-induced perturbation the same in all cases. **Figures 9A and 9D** represent two extreme scenarios; a strong Marangoni effect with negligible evaporation (**Figure 9A**) and a strong evaporation effect with negligible Marangoni effect (**Figure 9D**). In **Figure 9A**, the dominant Marangoni effect diminishes the surface concentration differences, resulting in a relatively uniform protein distribution on the surface. In contrast, when evaporation prevails (**Figure 9D**), our simulations show a distinct surface pattern due to a larger concentration difference. **Figures 9B and 9C** depict two competing scenarios where the Marangoni and evaporation effects are comparable in magnitude. In these cases, evaporation increases surface concentration differences, while Marangoni convection simultaneously weakens them, resulting in dynamically evolving patterns over time. As the effect of evaporation on the system increases (**Figure 9A→9C→9B→9D**), the time to form a surface pattern decreases.

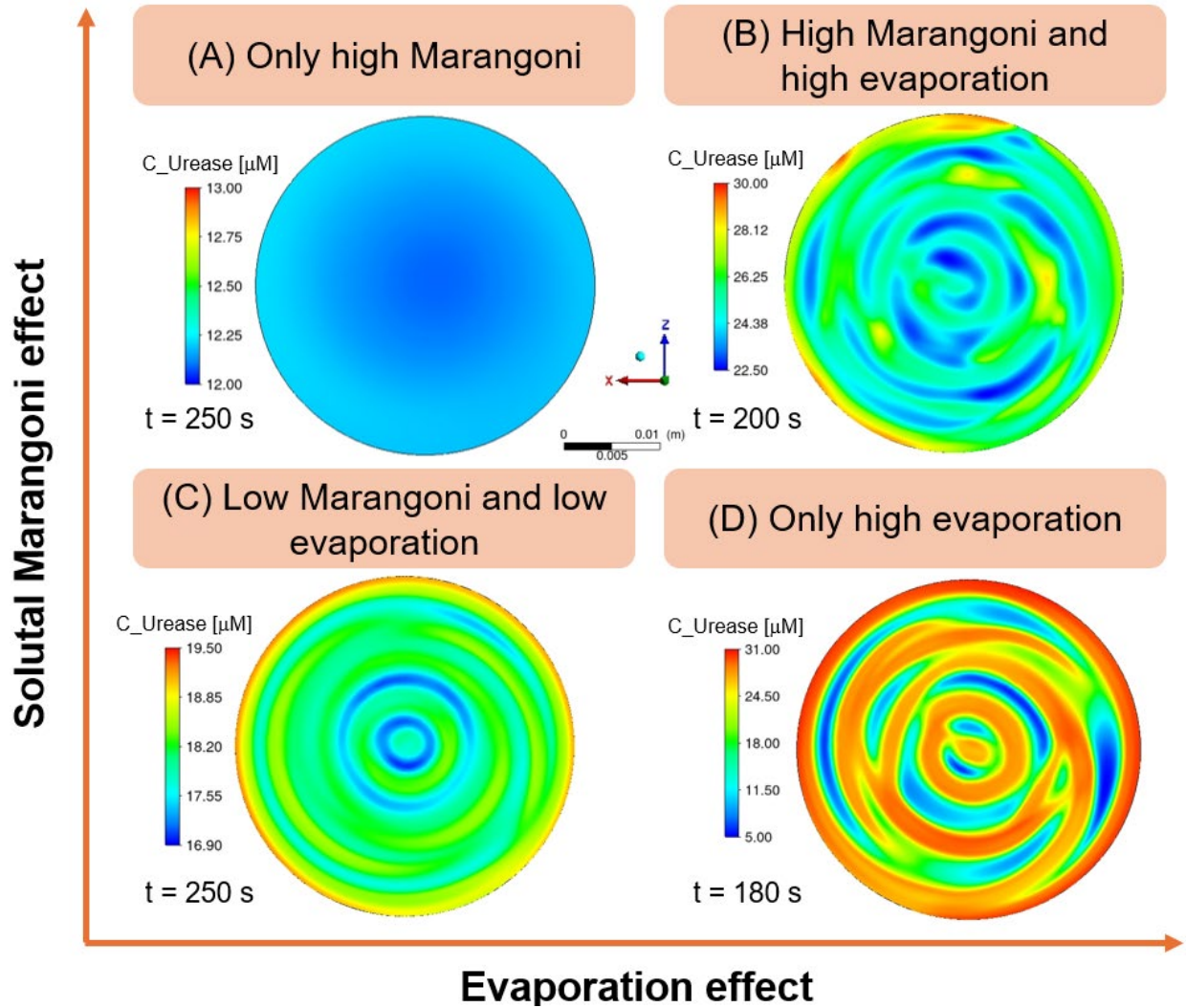

**Figure 9.** Patterns resulting from the interplay between evaporation and Marangoni effects. (A) In scenarios where the evaporation effect is negligible, the Marangoni effect predominates, leading to a relative uniform surface urease concentration (no pattern formed). (B) and (C) These two diagrams illustrate the competition between Marangoni and evaporation effects. In (C), the Marangoni gradient and evaporation rate intensities are halved compared to (B), highlighting the impact of reduced strengths on pattern dynamics. (D) When the Marangoni effect is negligible, patterns emerge primarily from high evaporative convection. In this condition, the absence of the Marangoni effect allows for a higher surface concentration difference of protein and a quicker pattern formation. For all four cases, a constant shear stress induced by airflow was maintained, while the strengths of evaporation and Marangoni effects were adjusted to reveal different pattern formations. Different contour legends are used in each panel in **Figure 9** to highlight urease local concentration distributions of the patterns.

## Conclusion

Through a combination of experiments and modeling, we have demonstrated novel pattern formation in an isothermal miscible fluid system involving simple protein and sugar solutions. Spiral patterns form over extended time scales due to the complex interplay of Marangoni, evaporation, and airflow effects. This finding is unexpected as solute Marangoni-related processes are generally characterized by fast spreading (seconds)[26,30,31], while the pattern formation in our systems take several minutes to form. Modeling suggests that this arises from different time scales for Marangoni and evaporative fluid flows. Our work suggests that Turing reaction-diffusion systems which are the mainstays in research on pattern formation[2,35] can be replicated by replacing the reaction-induced inhomogeneous solute distribution by evaporation-induced inhomogeneity. In both cases, the fast diffusive or Marangoni spreading of the solute is counteracted by a slower step that serves to reverse the solute homogenization. Our findings significantly expand the conditions that lead to pattern formation, demonstrating that dissipative patterns can form in the absence of thermal gradients or chemical reactions. The insights gained also enhance our ability to manipulate and control fluid motion and surface morphology, with promising implications for many areas such as coating technologies, materials science, and microfluidics.

## Materials and Methods

**Materials.** MES [2-(N-Morpholino)ethanesulfonic acid], glucose, sucrose, and Amicon Ultra centrifugal filter (10kDa), were purchased from Sigma Aldrich. Fructose was purchased from Alfa Aesar. Urease was purchased from TCI Chemicals. Dylight™ 550 maleimide was purchased from Thermo Fisher Scientific. Corning® 35 mm Not TC-treated Culture Dish was purchased from Neta Scientific Inc. Syringe (NORM-JECT ®R-F Luer Solo 1 ml) was purchased from Air-Tite Products Co., Inc. Needle (SOL-MTM Hypodermic Needle 27 G * 1/2'' (0.4 mm x 12.5 mm)) was purchased from Sol-Millennium Medical Inc. Tube (PTFE-28 Teflon Medical Tubing 0.015'' ID * 0.033" OD) was purchased from SAI Infusion Technologies.

**Dye labeling of protein.** To label urease, the protein and Dylight 550 maleimide were mixed in a molar ratio of 1:8 in 200 mM MES buffer (pH 6.0) and incubated overnight at room temperature with gentle rotation. The protein–dye conjugates were purified using pre-wetted Amicon Ultra centrifugal filter units (10 kDa molecular weight cutoff). The mixture was centrifuged at 4,500 rpm for 5 mins, and the retained conjugates were washed with buffer. This process was repeated twice. The concentration of the protein after purification was determined by measuring the absorbance at 280 nm using a Thermo Fisher UV-Visible Spectrophotometer. The protein concentration was adjusted before experiments.

**Pattern formation by proteins.** The schematic illustration of experimental setup is presented in **Figure S1**. A petri dish was filled with 3 mL of 200 mM MES buffer containing 100 mM of either D-glucose, sucrose, or MES buffer alone as a control. To each solution, 50 μL of 50 μM dye-tagged protein was introduced through a plastic tube. The additions were controlled by a syringe pump (CETONI, Base module 120, and NEMESYS low pressure module) and the flow speed of the pump is 1000 μL/min. The end of the tube was placed 0.5 cm from the solution surface in the petri dish. Videos were recorded using a camera at 30 frames per second for top view and side view.

In addition, the underwater fluid movement of pattern formation was also observed using a fluorescence microscope (Nikon ECLIPSE Ti-U). The focus plane was set to 0.5 mm below the air-water interface of the solution and Videos were recorded at 1 frame per second.

**Image analysis.** The pixel intensity profiles of pattern-formation videos were analyzed using ImageJ. At each time point, the intensity profiles along a line of interest were obtained and then subtracted by the intensity profile at 0 seconds, which was considered as the background intensity.

The movement of protein was obtained using Particle Image Velocimetry (PIV) with PIVlab from the MATLAB toolbox.[36,37] Prior to analysis, the videos were processed using CLAHE (contrast limited adaptive histogram equalization) to enhance contrast. The PIV analysis was conducted every second using FFT window deformation algorithm (direct Fourier transform correlation with multiple passes and deforming windows), with interrogation areas set to 64, 32, and 16 pixels. Vectors with low contrast were removed through a post-analysis filter. Finally, the average of velocity vectors at each location every 30 sec were calculated.

**Solution density measurements.** To determine the density of glucose and sucrose solutions, we measured the weight of 1 ml of each solution using a balance. Three independent measurements were conducted for each glucose or sucrose concentration. Linear regression was performed to identify the relationship between glucose/sucrose concentration and solution density.

**Surface tension measurements.** Different concentrations of urease solutions were prepared in 200 mM MES buffer with 100 mM glucose. The protein solutions were dispensed from a stainless-steel needle to form droplets. Surface tension measurements were obtained using a Ramé-hart 250-U1-R automatic goniometer. For each protein concentration, three independent measurements were conducted. During each measurement, the surface tension was recorded every three seconds over a twenty-minute period. For protein-containing solutions, the surface tension was calculated by averaging the readings between 11 and 15 min; for solutions without proteins, the average was calculated from measurements taken between 6 and 10 min.

**Numerical simulations for patten formation.** Computational fluid dynamics (CFD) models using Ansys Fluent software were developed to simulate the pattern formation for urease/glucose solution system. Species transport model was selected to represent the miscible fluid system, transient simulations were chosen to capture the dynamic changing of patterns. The laminar flow model was chosen given the low flow velocity observed during pattern formation stage, and gravity were activated to consider the buoyancy effect.

A 3D cylinder (diameter 3.5×10$^{-2}$ m, height 3.1×10$^{-3}$ m) (**Figure S13A**) was selected to represent the geometry of petri dish. Physical domain was discretized into small computational elements using Meshing tools, followed by a refinement for the near surface region (**Figure S13B**). A mesh independence study was conducted to ensure that the mesh was sufficiently refined. As illustrated in **Figure S13C**, the optimized number of mesh elements was 2,318,685, corresponding to a bulk mesh size of 3×10$^{-4}$ m and a surface mesh size of 1.5×10$^{-4}$ m. Further increases in the number of mesh elements were found to have no significant effect on the simulated Marangoni velocity. This setup achieves a balance between computational accuracy and efficiency.

The effects of Marangoni force, surface evaporation and airflow on pattern formation were investigated through simulation (See **Section S1** in **Supporting Information**). The Marangoni and airflow perturbation effects were modeled as shear stress conditions on the top surface boundary, which can be expressed as:

$$F_{shear} = F_M + F_{rotational}(r) \tag{Eq. 1}$$

$$F_M = \frac{\partial \sigma}{\partial C} * \nabla C \tag{Eq. 2}$$

Here, $F_M$ is the Marangoni force driving the spreading of the urease patch, determined by local surface tension gradient of the solution. The relationship between surface tension ($\sigma, [\frac{N}{m}]$) and protein concentration ($C, [\mu M]$) has been measured experimentally (**Figure S14**). Non-linear curve fitting ($R^2 = 0.99$) provided the following expressions of $\sigma(C)$ and $\frac{\partial \sigma}{\partial C}$

$$\sigma \left[\frac{N}{m}\right] = 10^{-3} * \left(52 + \frac{17.16}{1+\left(\frac{C}{0.89}\right)^{1.9}}\right) \tag{Eq. 3}$$

$$\frac{\partial \sigma}{\partial C}\left[\frac{N}{m*\mu M}\right] = 10^{-3} * \left(40.68 * \frac{C^{0.9}}{\left(1+\left(\frac{C}{0.89}\right)^{1.9}\right)^2}\right) \tag{Eq. 4}$$

The spatial concentration gradient $\nabla C$ in Eq. 2, which varies as the urease solution spreads on the surface, was determined during simulation by solving the mass and momentum governing equations.

As for the surface perturbation caused by airflow drag, we first confirmed the rotational flow type induced by airflow through experiments. Accordingly, a small shear stress term ($F_{rotational}(r)$) was set as the top surface boundary condition of our simulation model to induce surface motion. We then investigated the effects of symmetric and asymmetric surface shear stress on the resulting surface flow speed and pattern formation (**Figure S10**).

For surface evaporation, due to the difficulty in experimentally measuring local evaporation rates, the Hertz-Knudsen thermodynamic model was used to derive the evaporation flux function.[21] Assuming constant temperature and humidity, the evaporation rate was ultimately simplified as a function of protein concentration (see **Section S1** in **Supporting Information** for more details):

$$J_{water} \approx A_1 - A_2 * C_{urease} \qquad (Eq.\ 5)$$

Here, A1 [kg/(m²×s)] is the evaporation flux when protein concentration is zero, while A2 kg/(μM×m²×s) characterizes the protein concentration's effect on the evaporation rate. Referencing data from similarly scaled petri dish system[38–42] (**Table S2**), for demonstration purposes, we assumed an average water evaporation flux (A1) of 1×10⁻⁴ kg/(m²×s), with a concentration dependence A2 of 1×10⁻⁶ kg/(μM×m²×s). In Ansys Fluent, surface evaporation can be modeled as mass loss and momentum loss, and loaded into the first surface layer mesh via user-defined functions (define source and define initialization macro).[43]

To thoroughly investigate Marangoni spreading and pattern formation in this miscible fluid system, various simulation models were developed (e.g., different urease patch shapes and initial surface concentration distributions). A summary of all models used in this study is provided in **Table S1**. More information for governing equations (continuity, Navier-stoke momentum equation), material properties, meshing, model settings, and parameters can be found in **Section 1** in **Supporting Information**.

**Data and Code Availability**

Detailed information on simulation, including input parameters, conservation equations, settings, and algorithms, are provided in the Supporting Information. Simulation case files are available from A. Sen upon reasonable request.

**Acknowledgement**

We gratefully acknowledge financial support from the Sloan Foundation (Grant No. G-2023-19642). The authors acknowledge the Penn State Institute for Computational and Data Science for providing computational resources and support that have contributed to the research.

ASSOCIATED CONTENT

**Supporting Information.** Simulation settings and methods, supplementary figures and videos.

**Table S1.** Summary of types of models used in this research.

**Table S2**. Summary of reported water evaporation rates in different small systems.

**Figure S1**. Experimental setup.

**Figure S2**. Snapshots of pattern formation experiments. (side-view)

**Figure S3.** Densities of glucose and sucrose solutions.

**Figure S4**. Snapshots of pattern formation experiments with urease in 100 mM glucose.

**Figure S5.** Snapshots of pattern formation experiments observed by a fluorescence microscope.

**Figure S6**. Speed distribution of urease in the selected time interval observed by a fluorescence microscope.

**Figure S7**. Comparison of surface concentration and velocity vectors for two different starting urease patches during slow Marangoni spreading stage (t = 0.1s).

**Figure S8**. Local urease concentration distribution at 32 seconds.

**Figure S9**. Simulated irregular cell-like patterns with no airflow effects.

**Figure S10** Simulated spiral patterns with different surface shear stress models.

**Figure S11.** Snapshots of pattern evolution experiments with airflow perturbations.

**Figure S12**. Linear relationship between pattern formation time of urease and the density of glucose or sucrose solutions in the petri dish.

**Figure S13**: Physical geometry and Meshing details of petri dish model.

**Figure S14.** Relationship between urease concentration and corresponding solution surface tension.

**Video S1.** Dylight550-tagged urease in 200 mM MES buffer added to 100 mM glucose in 200 mM MES buffer. (10X speed)

**Video S2.** Dylight550-tagged urease added to 200 mM MES buffer only. (10X speed)

**Video S3.** Dylight550-tagged urease in 200 mM MES buffer added to 100 mM glucose in 200 mM MES buffer (Side-view). (10X speed)

**Video S4.** Dylight550-tagged urease added to 200 mM MES buffer only (Side-view). (10X speed)

**Video S5.** Dylight550-tagged urease suspended in 200 mM MES buffer with 100 mM glucose after being added in 200 mM MES buffer with 100 mM glucose. (10X speed)

**Video S6.** Dylight550-tagged urease suspended in 200 mM MES buffer with 100 mM glucose after being added to 200 mM MES buffer with 100 mM sucrose. (10X speed)

**Video S7.** Dylight550-tagged urease in 200 mM MES buffer added to 100 mM glucose in 200 mM MES buffer (fluorescence microscopy). (10X speed)

**Video S8.** Dylight550-tagged urease in 200 mM MES buffer added to 200 mM MES buffer only (fluorescence microscopy). (Speed up 10X)

**Video S9.** Simulation-Marangoni spreading (1.5X speed)

**Video S10**. Simulation-Irregular cell pattern (30~15X speed)

**Video S11**. Simulation-Spiral pattern (30~15X speed)

**Video S12**. Dylight550-tagged urease in 200 mM MES buffer added to 100 mM glucose in 200 mM MES buffer. A fan was used to create airflow above the solution surface. (10X speed)

**Video S13**. Dylight550-tagged urease in 200 mM MES buffer added to 100 mM glucose in 200 mM MES buffer. The petri dish was covered. (Speed up 10X speed)

**Video S14**. Adding 100 $\mu$M Dylight550-tagged urease in 200 mM MES buffer added to 100 mM glucose in 200 mM MES buffer. (Speed up 10X)